\documentclass[aps,pre,preprint, showpacs, superscriptaddress]{revtex4}
\usepackage[dvips]{graphicx}
\usepackage{latexsym}
\usepackage{natbib}       
\usepackage{epsfig}

\begin{document}
\title{Synchronization in Scale Free networks with degree correlation}
\author{C.~E.~La~Rocca} \affiliation{Instituto de Investigaciones
  F\'isicas de Mar del Plata (IFIMAR)-Departamento de F\'isica,
  Facultad de Ciencias Exactas y Naturales, Universidad Nacional de
  Mar del Plata-CONICET, Funes 3350, (7600) Mar del Plata, Argentina.}
\author{L. A. Braunstein} \affiliation{Instituto de Investigaciones
  F\'isicas de Mar del Plata (IFIMAR)-Departamento de F\'isica,
  Facultad de Ciencias Exactas y Naturales, Universidad Nacional de
  Mar del Plata-CONICET, Funes 3350, (7600) Mar del Plata, Argentina.}
\affiliation{Center for Polymer Studies, Boston University, Boston,
  Massachusetts 02215, USA} \author{P. A. Macri}
\affiliation{Instituto de Investigaciones F\'isicas de Mar del Plata
  (IFIMAR)-Departamento de F\'isica, Facultad de Ciencias Exactas y
  Naturales, Universidad Nacional de Mar del Plata-CONICET, Funes
  3350, (7600) Mar del Plata, Argentina.}
\begin{abstract}

In this paper we study a model of synchronization process on scale
free networks with degree-degree correlations. This model was already
studied on this kind of networks without correlations by Pastore y
Piontti {\it et al.}, Phys. Rev. E {\bf 76}, 046117 (2007). Here, we
study the effects of the degree-degree correlation on the behavior of
the load fluctuations $W_s$ in the steady state. We found that for
assortative networks there exist a specific correlation where the
system is optimal synchronized. In addition, we found that close to
this optimally value the fluctuations does not depend on the system size
and therefore the system becomes fully scalable. This result could be
very important for some technological applications. On the other hand,
far from the optimal correlation, $W_s$ scales logarithmically with
the system size.

\end{abstract}


\maketitle

\section{Introduction}

In the last decades the study of complex networks received much
attention because many real processes work over these kind of
structures. Historically, the research was mainly focused on how the
topology affects processes such as epidemic spreadings
\cite{Pastorras_PRL_2001}, traffic flow
\cite{Lopez_transport,zhenhua}, cascading failures \cite{Motter_prl}
and synchronization problems \cite{Jost-prl,Korniss07}. Many real
networks have structures characterized by a degree distribution $ P(k)
\sim k^{-\lambda}$ known as scale free (SF), where $k$ is the degree
or number of connections that a node can have and $k_{max} \ge k \ge
k_{min}$, where $k_{max}$ is the maximum degree, $k_{min}$ the minimum
degree and $\lambda$ measure the broadness of the distribution
\cite{Barabasi_sf}. In synchronization process it is customary to
study the fluctuations $W = \left\{1/N\sum_{i=1}^N (h_i-\langle h
\rangle)^2\right\}^{1/2}$ of some scalar field $h$, where $h_i$ with
$i=1,N$ represent the scalar field on node $i$, $\langle h \rangle$ is
the mean value, $N$ is the system size and $\{ . \}$ denotes an
average over network configurations. These kind of problems are very
important in many real situations such as supply-chain networks based
on electronic transactions \cite{Nagurney}, brain networks
\cite{JWScanell} and networks of coupled populations in correlated
epidemic outbreaks \cite{eubank_2004}. Pastore y Piontti {\it et. al}
\cite{anita} studied a model of surface relaxation with
non-conservative noise that allows to balance the load and reduce the
fluctuations (synchronize) of the scalar fields on SF networks without
degree correlation. However real networks are correlated in nature,
and there should be a reason for this feature. One reason could be to
enhance some process such as the transport and the synchronization
through them.  The degree-degree correlation of a network can be
measured using the Pearson's coefficient given by \cite{newman}
\begin{eqnarray}\label{pearson}
r=\frac{M^{-1}\sum_{i=1}^{M}j_ik_i-\left[M^{-1}\sum_{i=1}^{M}\frac{1}{2}(j_i+k_i)\right]^2}{M^{-1}\sum_{i=1}^{M}\frac{1}{2}(j_i^2+k_i^2)-\left[M^{-1}\sum_{i=1}^{M}\frac{1}{2}(j_i+k_i)\right]^2}\ ,
\end{eqnarray}
where $M$ is the number of edges of the network and $j_i$ and $k_i$
are the degree of the nodes of the edge $i$. This coefficient only can
takes values in the interval $[-1,1]$: if $r<0$ the network is called
disassortative (nodes with low degree tend to connect with highly
connected nodes) while for $r>0$ the network is called assortative
(nodes tend to connect with others with the similar degrees). When
$r=0$ the network is uncorrelated. As observed in many other works the
degree-degree correlation affects considerably the processes that
occur on top of them \cite{noh,ana2,sorrentino}.

In this paper we study the effects of the degree-degree correlation on
the behavior of the fluctuations in the steady state $W_s$ of SF
correlated networks with $\lambda < 3$ for the model of surface
relaxation to the minimum (SRM) \cite{family} used by Pastore y
Piontti {\it et. al} \cite{anita} in uncorrelated networks. To study
the fluctuations we map the process with a problem of a
non-equilibrium surface growth \cite{EW}, where the scalar field $h_i
\equiv h_i(t)$ represents the interface height at each node $i$ at
time $t$. We found that for every $\lambda < 3$ there exist a value of
the correlation for which the fluctuations are minimized, i.e, that
optimizes the synchronization. Close to and at the ``optimal''
correlation the fluctuations does not depend on $N$, but for other
correlations the fluctuations diverges logarithmically with $N$.

\section{Model and Simulation}

To construct the networks we use the configurational model (CM)
\cite{MolloyCorrelation} with a degree cutoff $k_{max}=N^{1/2}$ for
$\lambda < 3$ in order to uncorrelate the original network
\cite{correlacion}. Then, we choose two links at random and with
probability $p$ we connect the nodes with higher degree between them
and the two with smaller degree to each other to obtain $r>0$. For
$r<0$, we connect with probability $p$ the node with highest degree
with the one with lowest degree and the other two between them. In
both cases we do not allow self loops or multiple connections. It is
known that algorithms that generate clustering (the probability that
two connected nodes have another neighbor in common) produce
degree-degree correlation, but the algorithm used here produce
degree-degree correlation without introducing clustering
\cite{Newman-Miller}. In this way, we can study the effects of the
degree-degree correlation on SF networks isolating them from
clustering effects. A side effect of this algorithm is that for SF
networks the range of Pearson's coefficient that can be generated
cannot span the total domain $r \in [-1,1]$. Nevertheless, the range
that can be obtained is enough to observe how change the scaling of
the fluctuations with the system size when correlations are
introduced. For all the results in this work we use $k_{min}=2$ in
order to ensure that the network is fully connected \cite{cohen}. We
present the results for $\lambda=2.5$ but we checked that for $2 <
\lambda < 3$ they are qualitatively the same. The reason to
investigate only $2 < \lambda < 3$ is because almost all the real SF
networks fall in this range of values of $\lambda$.

In the SRM model \cite{anita,family}, at each time step a node $i$ is
chosen to evolve with probability $1/N$. Then, if we denote by $v_i$
the nearest neighbor nodes of $i$, the growing rules are: (1) if $h_i
\leq h_j$ $\forall j \in v_i$ $\Rightarrow h_i= h_i+1$, else (2) if
$h_j < h_n$ $\forall n \not= j \in v_i$ $\Rightarrow h_j = h_j+1$. For
the simulations we start with an initial configuration of $\{h_i\}$
randomly distributed in the interval $[0,1]$.

\begin{figure}[h]
\vspace{1cm}
\epsfig{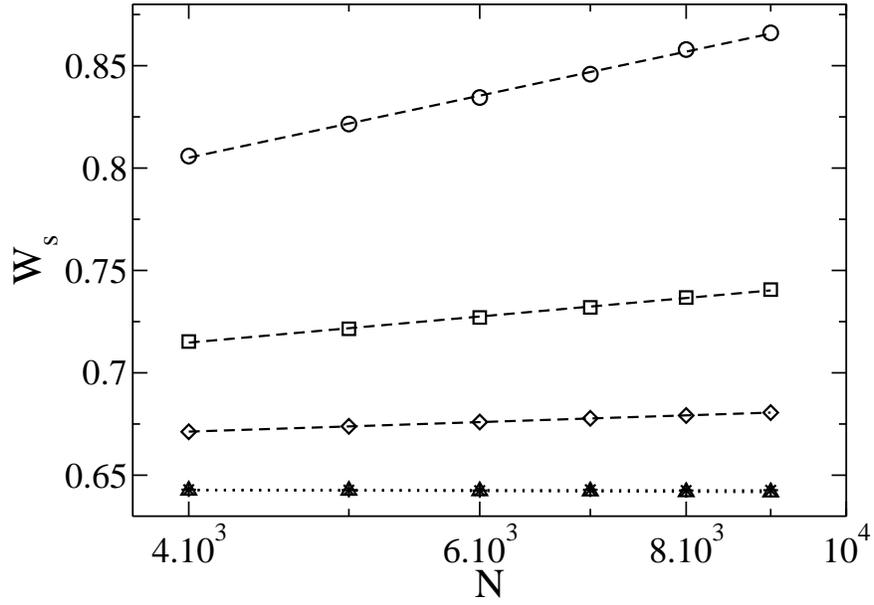}
%\hspace{1cm}
%\includegraphics[width=0.4\textwidth]{Fig1b.eps}
\caption{$W_s$ as a function of $N$ in log-linear scale for
  $\lambda=\ 2.5$ and $k_{min}=2$ for $r=-0.237$ ($\bigcirc$), $-0.05$
  ($\Box$), $0.103$ ($\diamond$), $r_{min}= 0.335$ ($\bigtriangleup$)
  and 0.386 ($\star$). The dashed lines represent a logarithmic
  fitting and the dotted lines a linear fitting.\\\label{fig.1}}
\end{figure}

\begin{figure}[h]
\vspace{1cm}
\epsfig{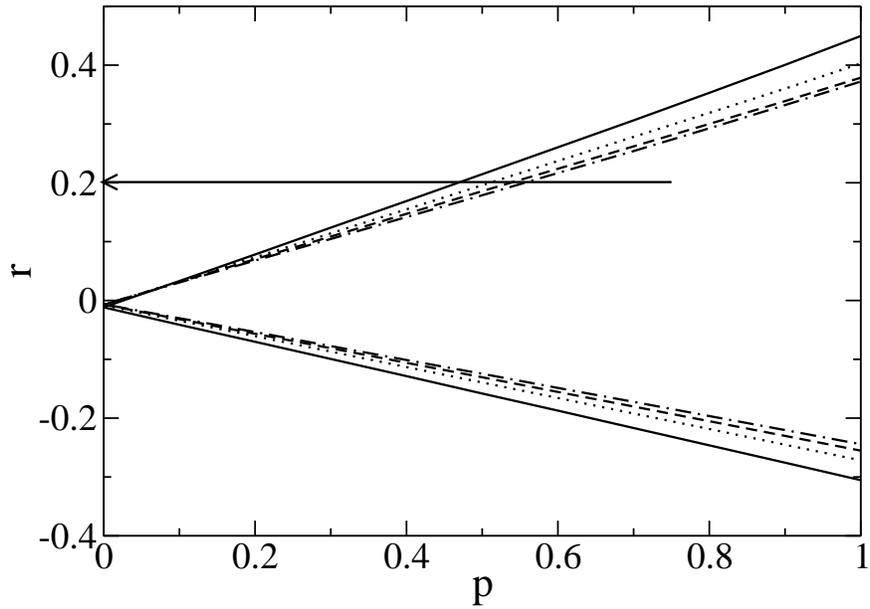}
%\hspace{1cm}
%\includegraphics[width=0.4\textwidth]{Fig1b.eps}
\caption{$r$ as a function of $p$ for $\lambda=\ 2.5$ and $k_{min}=2$
  for $N=1000$ (straight line), $3000$ (dotted line), $5000$ (dashed
  line) and $7000$ (dot-dashed line). The straight horizontal line
  with the arrow indicate the values of $p$ used for
  $r=0.2$. \\\label{fig.2}}
\end{figure}

In Fig.~\ref{fig.1} we show, in log-linear scale, $W_s$ as a function
of $N$ for different values of $r$ for $\lambda =2.5$. We can see that
for some values of $r$, $W_s$ has a logarithmic divergence with $N$
while for other values of $r$, $W_s$ does not depend or has a weakly
dependence on $N$. This change of behavior means that the scaling of
the fluctuations not only depends on $\lambda$ \cite{anita}, it also
depends on the correlation of the network. These results are in
agreement with Ref. \cite{anita}, where for uncorrelated, or slightly
disassortative networks, $W_s$ scales as $\ln N$ for $\lambda < 3$
($r=-0.05$ in Fig.~\ref{fig.1}). Notice that the relation between $r$
and $p$ has finite size effects (See Fig.~\ref{fig.2}). For this
reason if we want to fix $r$ we must select different values of $p$
for each system size.

\begin{figure}
\epsfig{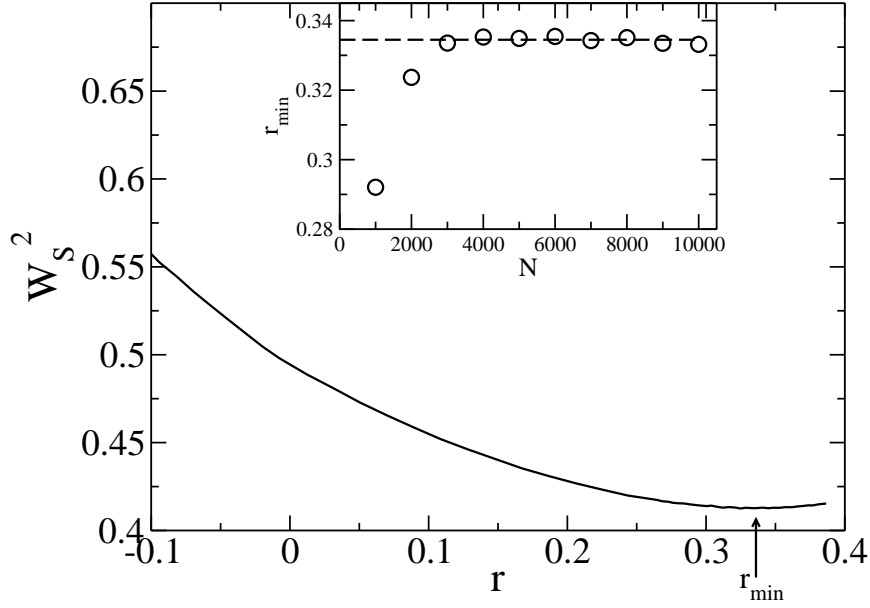}
\caption{$W_s^2$ as a function of $r$ for $\lambda=2.5$ and $N=5000$.
  The arrow indicate the position $r_{min}$. In the inset figures we
  plot $r_{min}$ as a function of $N$ in symbols. The dashed line
  represent the linear fitting in the region where $r_{min}$ is
  independent of $N$: $r_{min}=0.335$ for $\lambda=2.5$. The averages
  were done over $3000$ realizations.\\\label{fig.3}}
\end{figure}

In Fig.~\ref{fig.3} we plot $W_s^2$ as a function of $r$ for
$N=5000$. Each data point was obtained from the linear fitting of
$W^2(t)$ in the saturated regime for each $r$ value for $3000$
realizations, a task of very time consuming. We can see that there is
a positive value $r=r_{min}$ that minimizes (optimizes) the
fluctuations. In the inset figure we show $r_{min}$ as a function of
$N$. We can see that for large system size ($N \gtrsim 3000$) the
optimal correlation is independent of $N$. The dashed line represent
the linear fitting of $r_{min}$ for large $N$, from where we found
that $r_{min} \approx 0.335$ for $\lambda=2.5$. This means that for
the optimal correlation the fluctuations in the steady state do not
depend on the system size. This is an important result because for $r$
close to $r_{min}$ the whole system is scalable with $N$. As an
example, suppose that we have a cluster of computers connected as a SF
network with $r \approx r_{min}$, and that the excess of load of the
cluster of computers is sent to the first neighbors in one time step
as in our model. In the optimal correlation, as we show, the
fluctuations are independent of $N$, so we could increase the number
of computers in our system as much as we want without losing its
synchronization.

\begin{figure}[h]
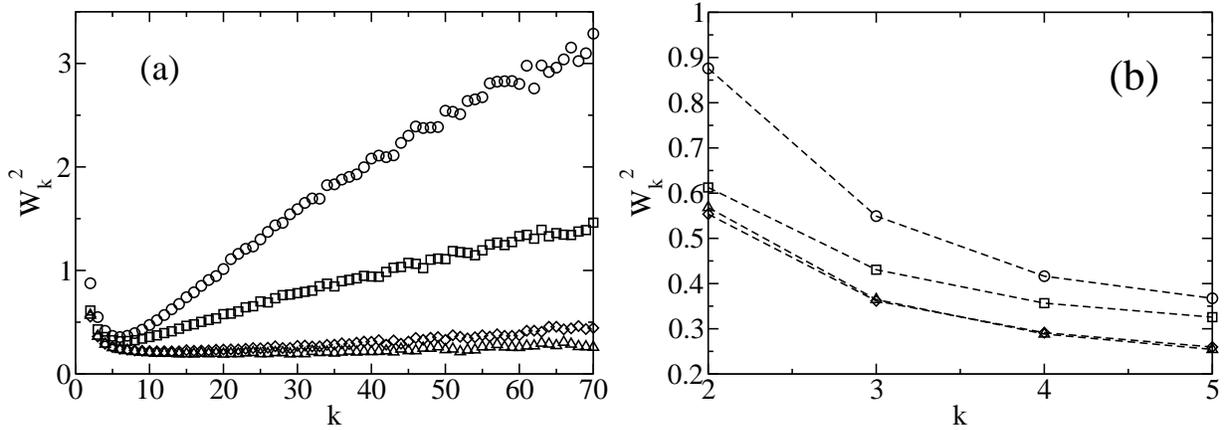

\epsfig{file=Fig4a.eps,width=8cm,angle=0}
\epsfig{file=Fig4b.eps,width=8cm,angle=0}
\caption{$W_k^2$ vs $k$ for $\lambda=\ 2.5$, $r=-0.255$ ($\circ$),
  $-7.10^{-03}$ ($\Box$), $0.335$ ($r_{min}$) ($\diamond$) and $0.386$
  ($\bigtriangleup$) b) An amplification of a) for low degree
  nodes. The dashed lines are used as guides. All these results are
  for $N=5000$ and $5000$ realizations of the
  networks.\\\label{fig.4}}
\end{figure}

In order to explain this behavior we compute the local contribution to
the fluctuations due to all nodes with degree $k$ in the steady state,
given by
\[W_k^2=\frac{1}{NP(k)}\sum_{i=1,k_i=k}^{N} W_i^2 \ ,\]
and therefore the total fluctuation can be computed as
\begin{equation}\label{rugosidad}
 W_s^2= \sum_{k=k_{min}}^{k_{max}} P(k)W_k^2 \ .
\end{equation}
In Fig.~\ref{fig.4} a) we show $W_k^2$ as a function of $k$ for
$\lambda=$ $2.5$, $N=5000$ and different values of $r$. We can see
that for nodes with high degree, $W_k^2$ decreases as $r$
increases. This is due to the fact that when $r$ increases nodes tend
to connect with others with similar connectivities and since the high
degree nodes are few and tightly packed, the one step relaxation is
enough for even out all their heights, balancing better the load and
enhancing the synchronization. On the other hand, for low degree nodes
we observe that $W_k^2$ has a minimum for the optimal correlation, as
shown in Fig.~\ref{fig.4} b). In SF networks the majority of the nodes
have low connectivities and if $r >0$ the average distance between
those nodes becomes bigger that one \cite{noh}. As the relaxation is
only to first neighbors, different parts of those chains will have
very different heights. As a consequence, for some values of $r>0$ not
all the low degree nodes will be completely synchronized between
them. For this reason the optimal correlation is positive and smaller
than one since low degree nodes are connected to some nodes with high
degree allowing to speed up the relaxation and smoothing out the
interface among them.

\begin{figure}
\epsfig{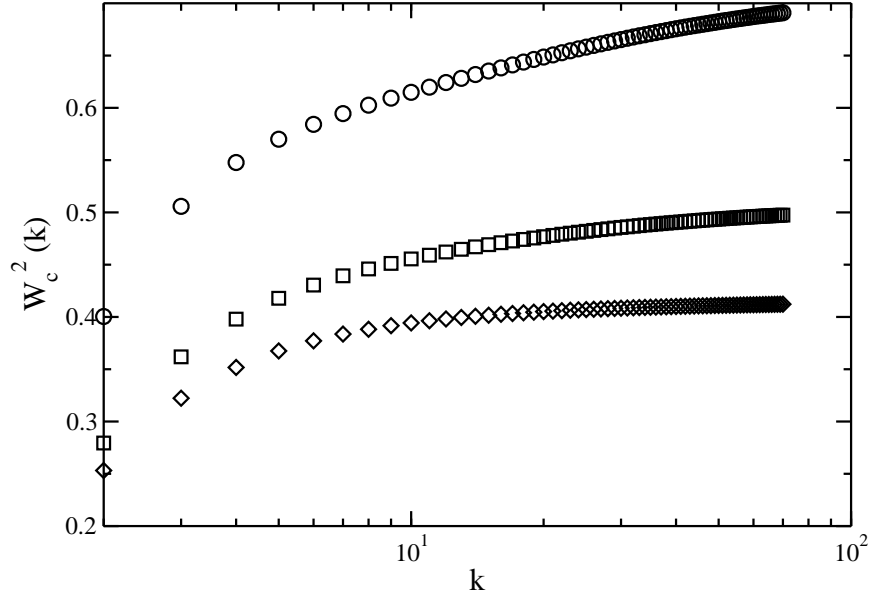}
\caption{Log-linear plot of $W_c^2(k)$ as a function of $k$ for
  $\lambda=\ 2.5$, $r=-0.255$ ($\circ$), $-7.10^{-03}$ ($\Box$) and
  $0.335$ ($r_{min}$) ($\diamond$). We can see that for $k > k^*$,
  where $k^* \approx 10$ the asymptotic behavior of $W_c^2(k)$ goes as
  $\ln k$ for $r < r_{min}$ and $r > r_{min}$ and goes as a $\mbox{const.}$ for
  $r \simeq r_{min}$. \\\label{fig.5}}
\end{figure}

In order to prove this, in Fig.~\ref{fig.5} we plot the cumulative
\begin{equation}\label{Wck}
 W_c^2(k)= \sum_{k^{'}=kmin}^k W_{k^{'}}^2 P(k^{'}) \ .
\end{equation}
As we can see from the plot, as $r$ increases the contribution to
$W_s$ of high degree nodes decreases, being the main contribution to
the fluctuations due to low degree nodes for positive $r$ and the
smaller for $r_{min}$. This is why the global fluctuations is minimal
in the optimal correlation. Also from the same plot we can understand
why close to the optimal correlation the system does not depend on
$N$. For $k > k^*$, Eq.~(\ref{Wck}) can be rewritten as
\[  W_c^2(k)= W_c^2(k^*) +  \sum_{k^{'}={k^{*}+1}}^{k} W_{k^{'}}^2 P(k^{'}) \ , \]
where $k^{*}$ is the upper value of $k$ that separate two different
regimes for $W_c^2(k)$. The second term can be replaced by $A(k)$,
where
\begin{eqnarray}\label{Ak}
A(k)  \sim \left \{
  \begin{array}{ll}
    \ln k , & \hbox{$r < r_{min}$ and $r > r_{min}$\ ;} \\
    \mbox{const.} , & \hbox{$r \simeq r_{min}$\ .}
  \end{array}
\right.
\end{eqnarray}
Then Eq.~(\ref{rugosidad}) is given by
\[ W_s^{2}= W_c^2(k_{max}) \ , \]
where $k_{max}=\sqrt{N}$. Using Eq.~(\ref{Ak})
\begin{eqnarray}\label{ws2}
W_s^2  \sim \left \{
  \begin{array}{ll}
    \ln k_{max} \sim \ln \sqrt N \sim \ln N , & \hbox{$r < r_{min}$ and $r > r_{min}$\ ;} \\
    \mbox{const.} , & \hbox{$r \approx r_{min}$\ .}
  \end{array}
\right.
\end{eqnarray}

The logarithmic divergence far from the optimal correlation is a
consequence of that the contribution of high degree nodes cannot be
disregarded. However in $r=r_{min}$ only the low degree nodes
contribute to the global fluctuations of the system allowing the
scalability with the system size.

\section{Summary}
In this paper we study the effects of degree degree correlations on
the behavior of the fluctuations $W_s$ for the SRM model in SF
networks with $ 2 < \lambda < 3$.  We found that there exist an
optimal value of the Pearson's coefficient $0 < r_{min} < 1$
(assortative networks) where the system is optimal synchronized. We
also found that for values close to $r_{min}$ the fluctuations does
not depend on $N$, \textit{i.e}, it is scalable. Moreover for $r <
r_{min}$ and $r > r_{min}$ the fluctuations diverge as a
logarithmically with the system size $N$.  Then the scaling behavior
of $W_s$ with the system size depend strongly on the correlation of
the network.

The optimal synchronization found for assortative networks is in our
model a topological effect due to the correlations. This is an
unexpected result because in many researches it was found that
disassortative networks (communication networks) are better for
transport \cite{ana2} and to synchronize oscillators
\cite{sorrentino}.

%and $0.386$ ($\bigtriangleup$)

\begin{thebibliography}{99}
\bibitem{Pastorras_PRL_2001} R. Pastor-Satorras and A. Vespignani,
  Phys. Rev. Lett. 86, 3200(2001).
\bibitem{Lopez_transport} E. L\'opez {\it et al.}, Phys. Rev. Lett. {\bf
94}, 248701 (2005); A. Barrat, M. Barth\'elemy, R. Pastor-Satorras and
A. Vespignani, PNAS {\bf 101}, 3747~(2004).
\bibitem{zhenhua} Z.~Wu, {\it et al.},  Phys.~Rev.~E.~\textbf{71},
045101(R)~(2005).
\bibitem{Motter_prl} A. E. Motter, Phys. Rev. Lett {\bf 93}, 098701
  (2004).
\bibitem{Jost-prl} J. Jost and M. P. Joy, Phys. Rev. E {\bf 65},
  016201 (2001); X. F. Wang, Int. J. Bifurcation Chaos
  Appl. Sci. Eng. {\bf 12}, 885 (2002); M. Barahona and L. M. Pecora,
  Phys. Rev. Lett. {\bf 89}, 054101 (2002); S. Jalan and
  R. E. Amritkar, Phys. Rev. Lett. {\bf 90}, 014101 (2003);
  T. Nishikawa {\it et al.}, Phys. Rev. Lett. {\bf 91}, 014101 (2003);
  A. E. Motter {\it et al.}, Europhys. Lett. {\bf 69}, 334 (2005);
  A. E. Motter {\it et al.}, Phys. Rev. E {\bf 71}, 016116 (2005).
\bibitem{Korniss07}  G. Korniss, Phys. Rev. E {\bf 75}, 051121 (2007).
\bibitem{Barabasi_sf} R. Albert and A.-L. Barab\'{a}si, Rev. Mod. Phys.
 {\bf 74}, 47 (2002); S. Boccaletti, V. Latora, Y. Moreno, M. Chavez and
 D.-U. Hwang, Physics Report {\bf 424}, 175 (2006).
\bibitem{Nagurney} A. Nagurney, J. Cruz, J. Dong, and D. Zhang,
Eur. J. Oper. Res. {\bf 164}, 120 (2005).
\bibitem{JWScanell} J. W. Scannell {\it et al.}, Cereb. Cortex 9, 277 (1999).
\bibitem{eubank_2004} S. Eubank, H. Guclu, V. S. A. Kumar, M. Marathe,
A. Srinivasan, Z.  Toroczkai and N. Wang, Nature {\bf 429}, 180 (2004);
M. Kuperman and G. Abramson, Phys Rev Lett {\bf 86}, 2909 (2001).
\bibitem{anita} A. L. Pastore y Piontti, P. A. Macri and
  L. A. Braunstein, Phys. Rev. E {\bf 76}, 046117 (2007).
\bibitem{newman} M. E. J. Newman, Phys. Rev. E {\bf 67}, 026126 (2003).
\bibitem{noh} Jae Dong Noh, Phys. Rev. E {\bf 76}, 026116 (2007).
\bibitem{ana2} A. L. Pastore y Piontti, L. A. Braunstein and
  P. A. Macri, Physics Letters A {\bf 374}, 4658-4663 (2010).
\bibitem{sorrentino} M. di Bernardo, F. Garofalo and F. Sorrentino,
  Proceeding of the 44th IEEE Conference on Decision and Control, and
  the European Control Conference 2005, Seville, Spain, December
  12-15, 2005 (4616-4621); arXiv:cond-mat/0506236v3.
\bibitem{family} F. Family, J. Phys. A {\bf 19}, L441 (1986).
\bibitem{EW} S. F. Edwards and D. R. Wilkinson, Proc. R. Soc. London, Ser. A
  {\bf 381}, 17 (1982).
\bibitem{MolloyCorrelation} M. Molloy and B. Reed, Random Struct. Algorithms
{\bf 6}, 161 (1995); Combinatorics, Probab. Comput. {\bf 7}, 295 (1998).
\bibitem{correlacion} S. Maslov and K. Sneppen, Science {\bf 296}, 910
  (2002); R. Xulvi-Bruner and I. M. Sokolov, Phys. Rev. E {\bf 70},
  066102 (2004); R. Xulvi-Bruner and I. M. Sokolov, Acta Physica
  Polonica B {\bf 36}, 1431 (2005).
\bibitem{Newman-Miller} M. E. J. Newman, Phys. Rev. Let. {\bf 103},
  058701 (2009); J. C. Miller, Phys. Rev. E {\bf 80}, 020901 (2009).
\bibitem{cohen} R. Cohen, S. Havlin, and D. ben-Avraham 446. Chap. 4
  in "Handbook of graphs and networks", Eds. S. Bornholdt and
  H. G. Schuster, (Wiley-VCH, 2002).
\end{thebibliography}
\end{document}